\documentclass[%
 reprint,
nofootinbib,
 amsmath,amssymb,amsfonts
 aps,
]{revtex4-1}

\usepackage{natbib}

\usepackage{xcolor}

\usepackage{bbold}
\usepackage{dsfont}
\usepackage{nccmath}
\usepackage{url}

\DeclareFontFamily{OT1}{pzc}{}
\DeclareFontShape{OT1}{pzc}{m}{it}{<-> s * [1.10] pzcmi7t}{}
\DeclareMathAlphabet{\mathpzc}{OT1}{pzc}{m}{it}

\usepackage[titletoc,toc,title]{appendix}
\usepackage{graphicx}% Include figure files
\usepackage{dcolumn}% Align table columns on decimal point
\usepackage{bm}% bold math
\usepackage[utf8]{inputenc}
\usepackage{amsmath}
\usepackage{amsfonts}
\usepackage{mathtools}
\usepackage{dsfont}
\usepackage{float} % Allows putting an [H] in \begin{figure} to specify the exact location of the figure
\usepackage{wrapfig} % Allows in-line images such as the example fish picture
\usepackage[caption=false]{subfig}

\usepackage[colorlinks=true,linkcolor=red]{hyperref} %Used for hyperlinking references within own paper and  for bib

\usepackage{cleveref} %Saves time, and auto puts in correct eq, fig, e.c.t

\usepackage{enumitem}

\usepackage{titlesec}

\titleformat{\paragraph}
{\normalfont\normalsize\bfseries}{\theparagraph}{1em}{}
\titlespacing*{\paragraph}
{0pt}{3.25ex plus 1ex minus .2ex}{1.5ex plus .2ex}

\begin{document}

%\preprint{APS/123-QED}

\title{Inflation and Scale-invariant $R^2$--Gravity}% Force line breaks with \\

\author{Carsten van de Bruck and Richard Daniel}
\affiliation{%
School of Mathematics and Statistics, University of Sheffield, Hounsfield Road, Sheffield S3 7RH, United Kingdom
}%

\date{\today}% It is always \today, today,
\begin{abstract}
In scale--invariant models of fundamental physics, mass scales are generated by spontaneous symmetry breaking. In this work, we study inflation in scale-invariant $R^2$ gravity, in which the Planck mass is generated by a scalar field, which is responsible for spontaneous breaking of scale--symmetry. If the self--interactions of the scalar field are non-zero, a cosmological constant is generated, which can be potentially quite large. To avoid fine--tuning at late times, we introduce another scalar field which drives the classical cosmological constant to zero during inflation. Working in the Einstein--frame, we find that due to a conserved Noether current the corresponding three--field inflationary model (consisting of the two scalar fields plus the scalaron) becomes effectively a two--field model. The prize to be paid for introducing the field which cancels the classical cosmological constant at the end of inflation is that the running of the spectral index and the running of the running can be quite large due to entropy perturbations during inflation, making the model testable with future cosmological experiments. 
%\begin{description}
%\end{description}
\end{abstract}

\pacs{Valid PACS appear here}% PACS, the Physics and Astronomy
\maketitle
\onecolumngrid
%\tableofcontents

\section{\label{sec:int}Introduction}
Cosmological inflation, a short period of accelerated expansion in the very early universe, is a simple and powerful idea \cite{Guth:1980zm,Starobinsky:1980te,Albrecht:1982wi,Linde:1981mu}. It provides an explanation for the current homogeneity and isotropy of our universe and suggests that the origins of the observed structures are results from quantum processes in the very early universe \cite{Mukhanov:1981xt,Hawking:1982cz,Bardeen:1983qw,Liddle:2000cg,Baumann_Inflation,Riotto_Inflation}. Predictions from inflation, such as the spectral index $n_s$ and the tensor-to-scalar ratio, fit observations such as those from the Planck experiment very well \cite{Planck2018_inflation,Chowdhury:2019otk}. Nevertheless, despite its appeal and simplicity, embedding inflation in fundamental theories of physics remains a challenge. On top of this, current observations indicate that the universe undergoes an epoch of accelerated expansion at the present time too. Both energy scales differ by many orders of magnitude. Finding an explanation from particle physics or theories of gravity for each epoch of accelerated expansion remains an active field of research. 

Inflation might be driven due to new matter sectors in a fundamental theory, or due to modifications of the gravitational sector or both \cite{Martin:2013tda}. Among the popular inflationary models, Starobinsky's original model of inflation, based on an additional $R^2$--term in the gravitational sector \cite{Starobinsky:1980te}, and its variants are promising models. There is considerable hope that modifications to General Relativity (GR) appear naturally in fundamental theories. From the phenomenological side, there have been many works studying the $R^2$--model and its extensions, see \cite{Edery:2019bsh,Gottlober:1990um,Rinaldi:2015uvu,Tambalo:2016eqr,Bamba:2015uxa,Antoniadis_single_R2,Calmet:f(r)_higgs,Capozziello:_f(r)_theories,DeFelice:f(R)_review,Tang:R2_higgs_DM,vandeBruck:r^2_extention,Ferreira_scale_independent_2019}. In this paper we consider an extension of $R^2$--gravity, in which there is no intrinsic mass scale, i.e. it is scale--invariant. The Planck mass, specified by the vacuum expectation value of a fundamental scalar field, is dynamically generated during inflation, similar to models studied in \cite{Rinaldi:2015uvu,Tambalo:2016eqr,Ferreira_scale_independent_2016,Ferreira_nofifthforces,Ferreira_scale_independent_2019,Kubo:2018kho,Kubo:2020fdd}. As it was shown in earlier work \cite{Rinaldi:2015uvu,Tambalo:2016eqr,Ferreira_scale_independent_2019}, and as we will review in Section 2, the prediction of the model for the spectral index and the tensor-to-scalar ratio are equivalent to the original Starobinsky model. However, as we will emphasise in Section 2, a classical cosmological constant is generated after inflation, which can be potentially large, unless the self--interaction of the scalar field is unnaturally small or zero if the classical cosmological constant is tuned to be zero. In this paper we extend the model by introducing another scalar field which dynamically drives the cosmological constant to zero during inflation. As we will show, the dynamics of this second field will generate features in the primordial power spectrum and a cut-off at small wavelength. After inflation, the theory is well approximated by GR with vanishing classical cosmological constant. The scalaron field, describing modifications from GR due to the $R^2$ term, is heavy and has therefore a very small interaction range. 

The paper is organized as follows. In Section II we discuss the prediction of scale invariant $R^2$ inflation, which in the Einstein frame is a two--field model. We recover the results of previous work, highlighting the fact that the theory predicts a large cosmological constant in the Einstein frame unless the field determining the Planck mass has no self-interactions. In Section III we present a modification by adding another scalar field, whose role is to cancel the cosmological constant. In the Einstein frame, the theory is described by a three--field system. The inflationary dynamics and results from the numerical computations are discussed in Section IV. Our conclusions can be found in Section V. In the appendix we present the perturbation equations and describe an extension of \cite{Wands_2000_perturbations,Lalak_pertubations} in performing a field rotation into adiabatic and entropy fields for a three--field inflationary system and also discuss the numerical setup.  

\section{\label{sec:two}A two-field model}
The scale--independent models we study in this paper are extensions of Starobinsky's $R^2$-model \cite{Starobinsky:1980te}, in which there is no intrinsic mass scale. Instead, the Planck mass is generated dynamically during inflation. In this Section we will consider one additional scalar field $\chi$ to the $R^2$ model, which couples to the Ricci--scalar. The action is specified by 
\begin{equation}\label{two-field}
    {\cal S} = \int d^4x \sqrt{-\tilde{g}}\left( f(\tilde{R}, \chi) - \frac{1}{2} \tilde{g}^{\mu\nu} \partial_\mu \chi \partial_\nu \chi - V(\chi) \right),
\end{equation}
with 
\begin{eqnarray}
f(\tilde{R},\chi) &=& \frac{1}{2}A \chi^2 \tilde{R} + \frac{B^2}{2}\tilde{R}^2 \label{R2model1}\\
V(\chi) &=& \frac{\lambda}{4}\chi^4 \label{R2model2}.
\end{eqnarray}
Here, $A, B$ and $\lambda$ are constant parameter. This set-up has previously been studied in the context of inflation in \cite{Tambalo:2016eqr,Ferreira_scale_independent_2019}. We will recover some of the results of the literature in the following, emphasising the fact that the cosmological constant at the end of inflation is non-zero and potentially quite large unless $\lambda=0$ or unnaturally small.

To analyse the predictions from inflation, we will study the model in the Einstein--frame. Theories of the $f(R,\chi)$--type can be brought into the Einstein--frame via a conformal transformation of the metric (see \cite {Hwang:1996np, Rodrigues_Auxiliary_fields} and e.g. \cite{DeFelice:f(R)_review} and references therein). Defining the auxiliary field $\exp(\beta\kappa\psi)/2\kappa^2 = f_{,\tilde{R}}$ with $\beta = \sqrt{2/3}$ and considering the conformal transformation $\tilde{g}_{\mu\nu} = \exp(-2\beta\kappa\psi) g_{\mu\nu}$, the action can be brought into the Einstein frame, which takes the form 
\begin{equation}\label{ouraction}
    {\cal S} = \int dx^4 \sqrt{-g} \left[\frac{R}{2\kappa^2} - \frac{1}{2}g^{\mu\nu}\partial_\mu \psi \partial_\nu \psi - \frac{1}{2} e^{-\beta\kappa\psi}\partial_\mu \chi \partial_\nu \chi - V_{\rm T}(\psi,\chi) \right],
\end{equation}
where we identify $\kappa=M_{\rm Pl}^{-1}$. The total potential energy $V_{\rm T}(\psi,\chi)$ for our model will be specified below. Note that $\kappa$ sets an arbitrary mass-scale, which we have identified with the Planck mass in the Einstein frame. 

Theories of the form (\ref{ouraction}) have been considered in e.g.  \cite{Finelli_2field_perturbation,Lalak_pertubations,vandeBruck:2014ata,vandeBruck:r^2_extention,vandeBruck:2016rfv}. For the $R^2$--model we are considering in this paper, specified by equations (\ref{R2model1}) and (\ref{R2model2}), we have
\begin{equation} \label{eq:potential}
  4\kappa^4 V_{\rm T}(\psi,\chi) = \frac{1}{2B^2}\left(1 - A\kappa^2\chi^2 e^{-\beta\kappa\psi} \right)^2 + \lambda \chi^4 e^{-2\beta\kappa\psi}.
\end{equation}
At the minimum of the potential, specified by $V_{{\rm T},\psi} = 0 = V_{{\rm T},\chi}$, we find
\begin{eqnarray}
\frac{M_{\rm Pl}^2}{2}\frac{e^{\beta\kappa\psi_{\rm min}}}{\chi_{\rm min}^2} = \frac{B^2}{A}\left(\lambda + \frac{A^2}{2B^2} \right)
\end{eqnarray}
The potential energy at the minimum is given by 
\begin{eqnarray}
    V_{\rm T,min} = \frac{M_{\rm Pl}^4}{4} \frac{\lambda}{2B^2 \lambda + A^2}.
    \label{eq 2fieldmin}
\end{eqnarray}
Thus, as long as $\lambda$ does not vanish, a cosmological constant is generated\footnote{In the Jordan--frame, after inflation the vacuum expectation value of $\chi$ does not vanish. The potential energy will not vanish either if $\lambda$ is non-zero, resulting in a positive cosmological constant.}. Furthermore, we can calculate the mass of the scalaron $\psi$ at the minimum. We find 
\begin{equation}\label{eq:mass}
    \frac{m_\psi^2}{M_{\rm Pl}^2} = \frac{1}{6B^2}\frac{A^2}{2B^2\lambda + A^2}
\end{equation}
Therefore, in this model the cosmological constant and the scalaron mass are linked via
\begin{eqnarray}\label{eq, mass potential}
    V_{\rm T,min} = \frac{3}{2}\lambda \frac{B^2}{A^2}m_{\psi}^2M_{\rm Pl}^2.
\end{eqnarray}

The Jordan--frame action (\ref{two-field})-(\ref{R2model2}) is invariant under the global Weyl transformation 
\begin{eqnarray}
    g_{\mu\nu}&\rightarrow& e^{2\epsilon} g_{\mu\nu}\nonumber \\
    \chi &\rightarrow& e^{-\epsilon} \chi.
\end{eqnarray}
In the Einstein frame, with the action given by (\ref{ouraction}) and the potential energy given by (\ref{eq:potential}), there exists a corresponding symmetry as well: the field--transformation 
\begin{eqnarray}
    \chi &\rightarrow& e^\epsilon \chi \nonumber \\
    \psi &\rightarrow& \psi + \frac{\sqrt{6}}{\kappa^2}\epsilon,
\end{eqnarray}
leaves the Einstein--frame action invariant. As a result, the following current is conserved 
\begin{equation}
    j_\mu = \sqrt{-g} \partial_\mu \left( 3 e^{\beta\kappa\psi} + \frac{1}{2}\chi^2 \right)~,
\end{equation}
i.e. 
\begin{equation}\label{conservedcurrent}
    \partial_\alpha j^\alpha = 0
\end{equation}
Considering a flat Robertson--Walker metric with 
$$ds^2 = -dt^2 + a^2(t) \delta_{ij}dx^i dx^j,$$ 
where $a(t)$ is the scale factor, this equation implies that, in cosmology, 
\begin{equation}
    {a^3}\frac{d}{dt}\left[3 e^{\beta\kappa\psi} + \frac{1}{2}\chi^2 \right] = {\rm constant},
\end{equation}
Therefore, during inflation in which the scale factor grows quasi--exponentially, we quickly approach a regime where
\begin{equation}
    3e^{\beta\kappa\psi} + \frac{1}{2}\chi^2 = {\rm constant} \equiv 3\tilde{c},
\end{equation}
and the two--field model quickly becomes effectively a one--field system. 
This allows us to rewrite eq. \eqref{eq:potential} as 
\begin{equation}\label{eq, onefield potential}
    \kappa^4V_T={9}\left(1 - \tilde{c}e^{-\beta\kappa\psi} \right)^2  \left(\frac{A^2}{2B^2} +\lambda \right),
\end{equation}
which is the effective potential during inflation at sufficient late times. 

We now turn to find the constraints on the parameter of the model. Apart from constraints coming from the spectral index, the tensor-to-scalar ratio and the amplitude of scalar perturbations generated during inflation, we will discuss the implications of the non--vanishing vacuum energy density in the model too. 

To analyse the evolution of the fields during inflation, we state the slow-roll parameters of our model   
\begin{align}\label{epsilon}
    \epsilon_v= \frac{1}{2\kappa^2}\left(\frac{V_{T,\psi\psi}}{V}\right)^2=\frac{4}{3}\frac{{\tilde c}^2}{(e^{\beta\kappa\psi}-\tilde{c})^2},
\end{align}
\begin{equation}\label{eta}
    \eta_v=\frac{1}{\kappa^2} \frac{V_{T,\psi\psi}}{V} = -\frac{4}{3}{\tilde{c}}\frac{e^{\beta\kappa\psi}-2\tilde{c}}{\left(e^{\beta\kappa\psi}-\tilde{c} \right)^2}.
\end{equation}
We have a successful slow-roll inflationary period while $\epsilon<1$, which sets the field value at the end of inflation to be $\exp(\beta\kappa\psi_{end})=({\tilde{c}+\tilde{c}\sqrt{3}})/{\sqrt{3}}.$ From the slow roll parameters we determine the inflationary observables, the scalar spectral index, $n_s$ and the tensor-to-scalar ratio, $r$, from the background fields,
\begin{align}
    n_s \simeq 1-\frac{2}{N}, \hspace{1cm} r\simeq \frac{12}{N^2}.
\end{align}
Here we have written $n_s$ and $r$ in terms of the number of e-folds, $N$, to highlight the fact that the prediction for these observables are the same as in Starobinsky inflation. We will assume that the horizon crossing is at $N\approx 60$.
We can then finally begin to constrain the model parameters from the energy scale of inflation, $V\simeq100r \cdot(10^{16}\text{GeV})$. In our model we use the form of \eqref{eq, onefield potential} which gives
\begin{equation}
    V\simeq\left(\frac{A^2}{2B^2}+\lambda\right)=10^{-12}
\end{equation}

The interaction range of the scalaron is less than a millimeter, implying that $m_{\psi}^{-2}<1{\rm mm}^2$. This allows us to further constrain our model via \eqref{eq:mass} to give 
\begin{equation}
    \frac{B^4}{A^2} \lesssim 10^{77}.
\end{equation}
Finally, we demand that the energy density at the minimum drives the cosmological expansion today, i.e. we set $V_T=10^{-122}M_{{\rm Pl}}$ at the minimum to be the observed dark energy density. This gives the result
\begin{equation}
    A\simeq 10^{-6} B, \hspace{1cm} B\lesssim10^{33} \hspace{0.5cm}\text{and}\hspace{0.5cm} \lambda<10^{-65}.
\end{equation}
This result implies that $\lambda$ has to be unnaturally small, coming from the demand that the cosmological constant is small. Of course, dark energy might originate from a different sector in the theory and we could simply demand that $\lambda = 0$. On the other hand, rather than relying on the self--interactions to vanish or to be unnaturally small, it would be more satisfying if the (classical) cosmological constant is dynamically driven to zero during inflation. In the next section we discuss a model in which exactly this happens. To achieve this, we employ another scalar field. 

\section{\label{sec:three}A three--field model}
The action we consider is a variant of the action (\ref{two-field}), with the addition of a second field $\sigma$:
\begin{equation}\label{three-field}
    {\cal S} = \int d^4x \sqrt{-\tilde{g}}\left( f(\tilde{R}, \chi) - \frac{1}{2} \tilde{g}^{\mu\nu} \partial_\mu \chi \partial_\nu \chi - \frac{1}{2} \tilde{g}^{\mu\nu} \partial_\mu \sigma \partial_\nu \sigma - V(\chi,\sigma) \right),
\end{equation}
with 
\begin{eqnarray}
f(\tilde{R},\chi) &=& \frac{1}{2}A \chi^2 \tilde{R} + \frac{B^2}{2}\tilde{R}^2 \label{R2model3}\\
V(\chi,\sigma) &=& \frac{\lambda}{4} (\chi^2 - \sigma^2)^2. \label{R2model4}
\end{eqnarray}
Here we couple only $\chi$ to the Ricci--scalar and only $\chi$ determines the value of the Planck mass. An extension of the model in which also the $\sigma$--field couples to the Ricci--scalar is possible, but that would introduce another parameter. The role of the field $\sigma$ is to drive the cosmological constant in the Einstein--frame to zero. 

In the Einstein frame the action becomes a three-field system and reads 
\begin{equation}
    {\cal S} = \int d^4 x \sqrt{-g} \bigg\{ \frac{R}{2\kappa^2} - \frac{1}{2} g^{\mu\nu} \left[\partial_\mu\psi\partial_\nu\psi+e^{-\beta\kappa\psi}\left( \partial_{\mu} \chi \partial_{\nu} \chi + \partial_{\mu} \sigma \partial_{\nu} \sigma\right)\right]  - V_{\rm T}(\psi,\chi,\sigma) \bigg\},
\label{eq full_einstein}
\end{equation}
with the potential 
\begin{equation}
    V_{\rm T}(\psi,\chi,\sigma) = \frac{1}{8B\kappa^4}\left(1 - A\kappa^2\chi^2 e^{-\beta\kappa\psi} \right)^2 + \frac{\lambda}{4} \left(\chi^2-\sigma^2\right)^2 e^{-2\beta\kappa\psi}.
    \label{eq full_potential}
\end{equation}
With the addition of the new field, the potential has now a global minimum at which $V_{\rm T}=0$. This is similar to a dynamical Higgs VEV model shown in \cite{GarciaBellido_dynamical_higgs}. At the minimum we have 
\begin{equation}
    \chi_{\rm min}^2=\sigma_{\rm min}^2, \hspace{1cm} e^{\beta\kappa\psi_{\rm min}}=\kappa^2A\sigma_{\rm min}^2.
\end{equation}
As a consequence, $\lambda$ is now only constrained by the amplitude of primordial scalar perturbations. 

The equations of motions for the fields in our model are given by  
\begin{align}
\label{eom psi}
    \Box\psi&=\frac{\partial V_{\rm T}}{\partial\psi} - \frac{\beta\kappa}{2}e^{-\beta\kappa\psi} g^{\mu\nu}\left(\partial_\mu \chi \partial_\nu \chi + \partial_\mu \sigma \partial_\nu \sigma\right),\\
\label{eom chi}
    \Box\chi& - \beta\kappa g^{\mu\nu}\partial_\mu\psi \partial_\nu\chi = \frac{\partial V_{\rm T}}{\partial\chi}e^{-\beta\kappa\psi},\\
\label{eom g}
    \Box\sigma& - \beta\kappa g^{\mu\nu}\partial_\mu\psi \partial_\nu\sigma = \frac{\partial V_{\rm T}}{\partial\sigma}e^{-\beta\kappa\psi}, 
\end{align}
In an expanding homogeneous and isotropic universe, the Friedmann equation reads  
\begin{equation}
    H^2=\frac{\kappa^2}{3}\left[\frac{1}{2}\Dot{\psi}^2 + \frac{1}{2}e^{-\beta\kappa\psi}\left( \Dot{\chi}^2+ \Dot{\sigma}^2\right)+ V_{\rm T} (\psi,\chi,\sigma) \right].
\end{equation}
The action is invariant under the following transformations of the fields 
\begin{eqnarray}
    \sigma &\rightarrow& e^\epsilon \sigma, \nonumber \\
    \chi &\rightarrow& e^\epsilon \chi, \nonumber \\
    \psi &\rightarrow& \psi + \frac{\sqrt{6}}{\kappa^2}\epsilon, \nonumber 
\end{eqnarray}
and as before there is a conserved current, which allows us to reduce our system to an effective two field case at late times during inflation. Following a calculation similar to the two-field case, we find that at late times 
\begin{equation}
    \chi^2 + \sigma^2 = {\cal C} - \frac{6}{\kappa^2}e^{\beta\kappa\psi}, 
    \label{eq:noether2}
\end{equation}
where $\cal C$ is a constant of integration, specified by initial conditions for the fields. This has been confirmed numerically with various initial conditions and parameters. We will use this result in the next section to simplify and understand the results of our model. 
\section{Background evolution and cosmological perturbations} 
% Figures ------------------------------------------------------------------------------------------------------------------
\begin{figure}
    \centering
    \includegraphics[scale=0.7]{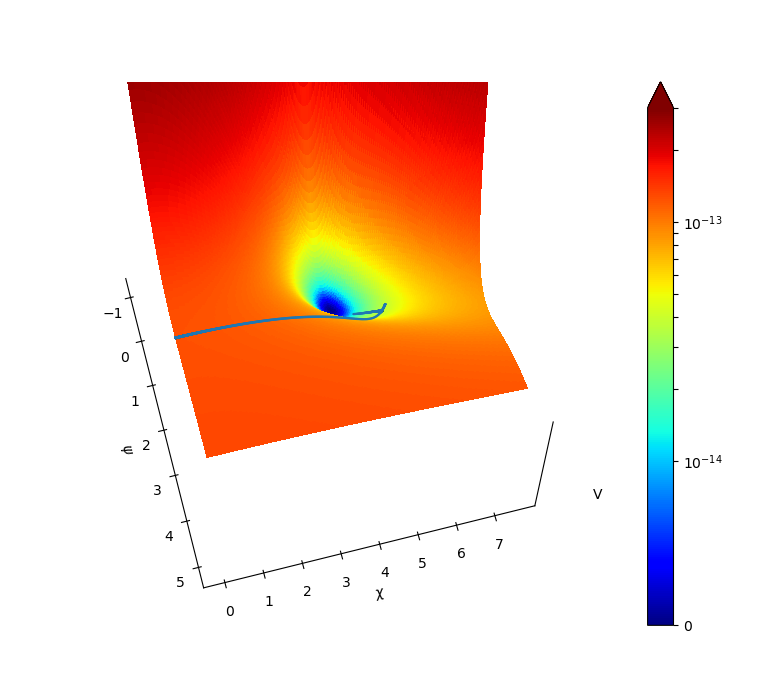}
    \caption{A 3D surface plot above its contour of the three field potential \eqref{eq full_potential}. Here we have used the surface plot to illustrate the two slow-roll regimes, with the trajectory plotted on top in blue. We have set, $A=0.05$, $B=10^6$, $\lambda=10^{-15}$ and used eq. \eqref{eq:noether2} to trade $\sigma$ in favour of $\chi$ and $\psi$ with ${\cal C} = 7.7$, to match the used initial conditions of \cref{fig background}. The colour bar indicates the value of the potential in Planck units. }
    \label{fig new potential}
\end{figure}

\begin{figure}
    \centering
    \includegraphics[scale=0.5]{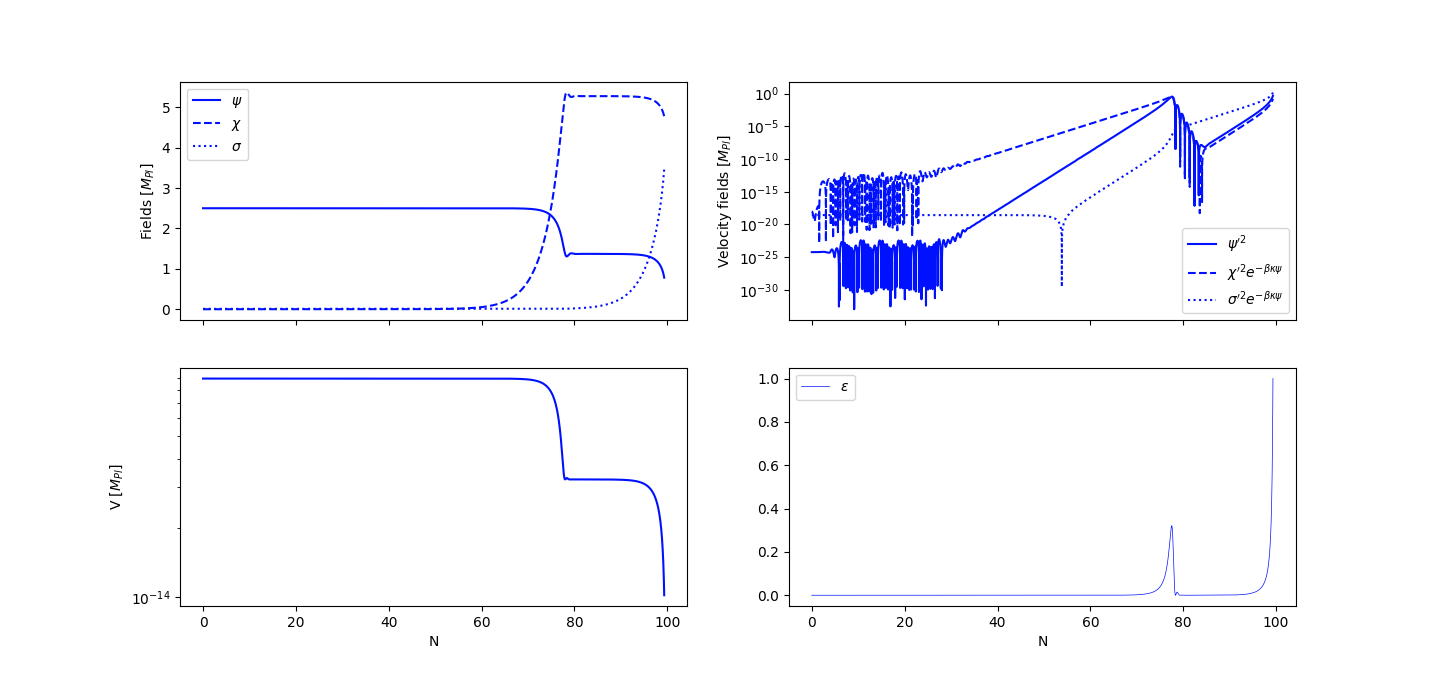}
    \includegraphics[scale=0.5]{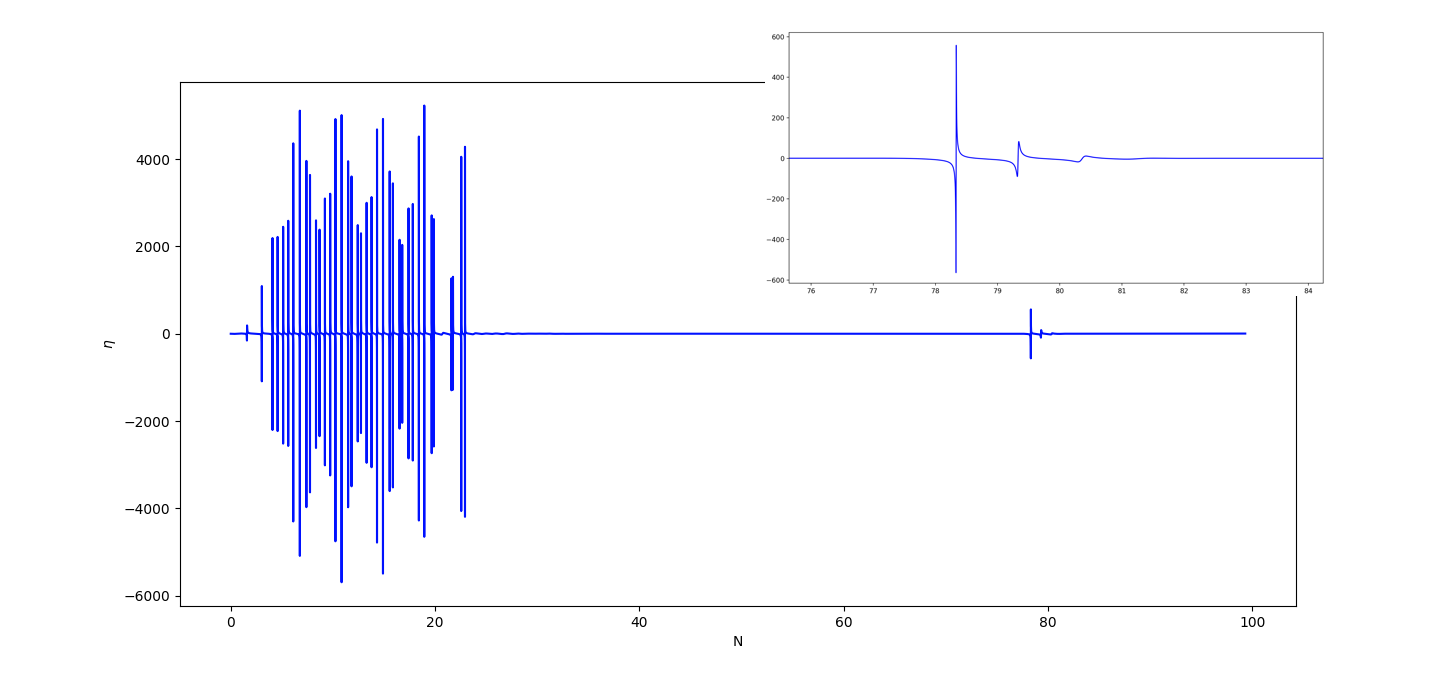}
    \caption{[Top left] The evolution of the fields, note the how $\psi$ and $\chi$ freeze in second period of inflation, [Top right] The velocity of the fields, [Middle left] A 2D plot of the potential clearly illustrating two periods of slow roll inflation, [Middle right] Slow roll parameter, $\epsilon$. 
    [Bottom] The slow roll parameter $\eta$, with additional zoomed in plot of late time slow roll violation.
    Here we have used the parameters and initial conditions: $A=0.05$, $B^2=2\times10^{12}$, $\lambda=10^{-15}$, $\psi= 2.5$, $\chi=10^{-8}$, and $\sigma= 10^{-2}$}
    \label{fig background}
\end{figure}

\begin{figure}
    \centering
    \advance\leftskip-2cm
    \advance\rightskip-2cm
    \includegraphics[width=0.5\textwidth]{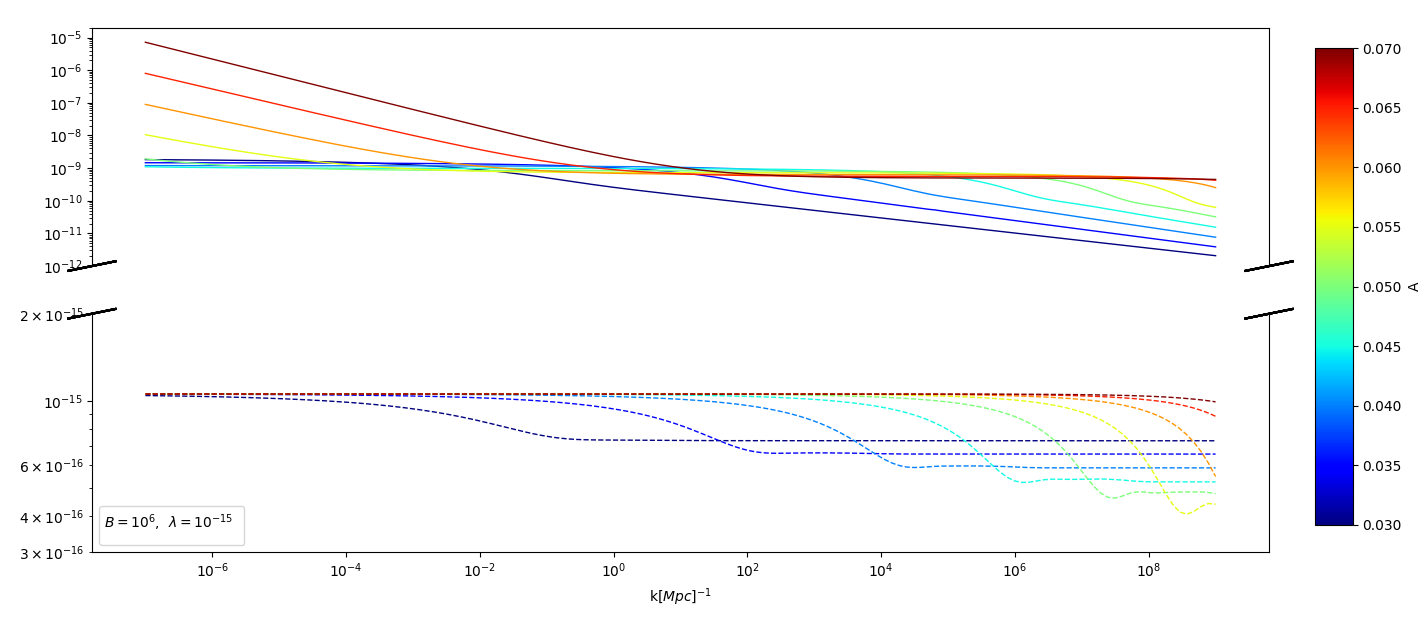}
    \includegraphics[width=0.5\textwidth]{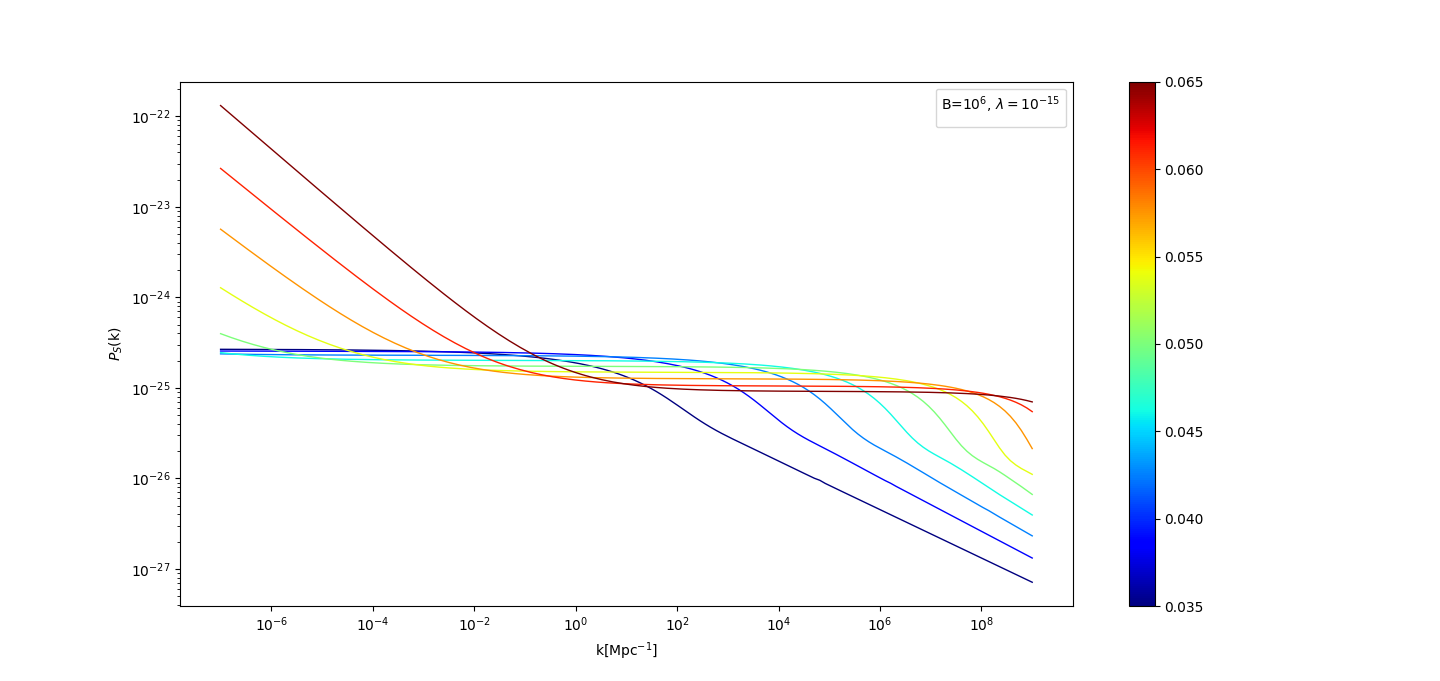}
    \includegraphics[width=0.5\textwidth]{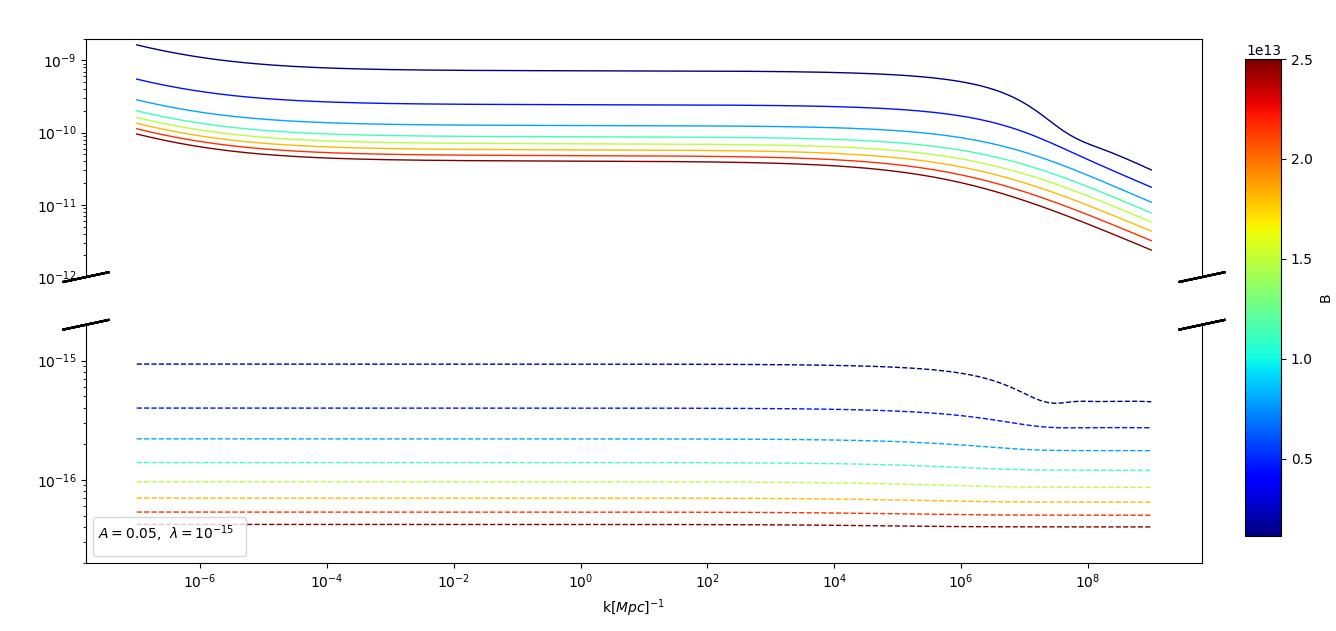}
    \includegraphics[width=0.5\textwidth]{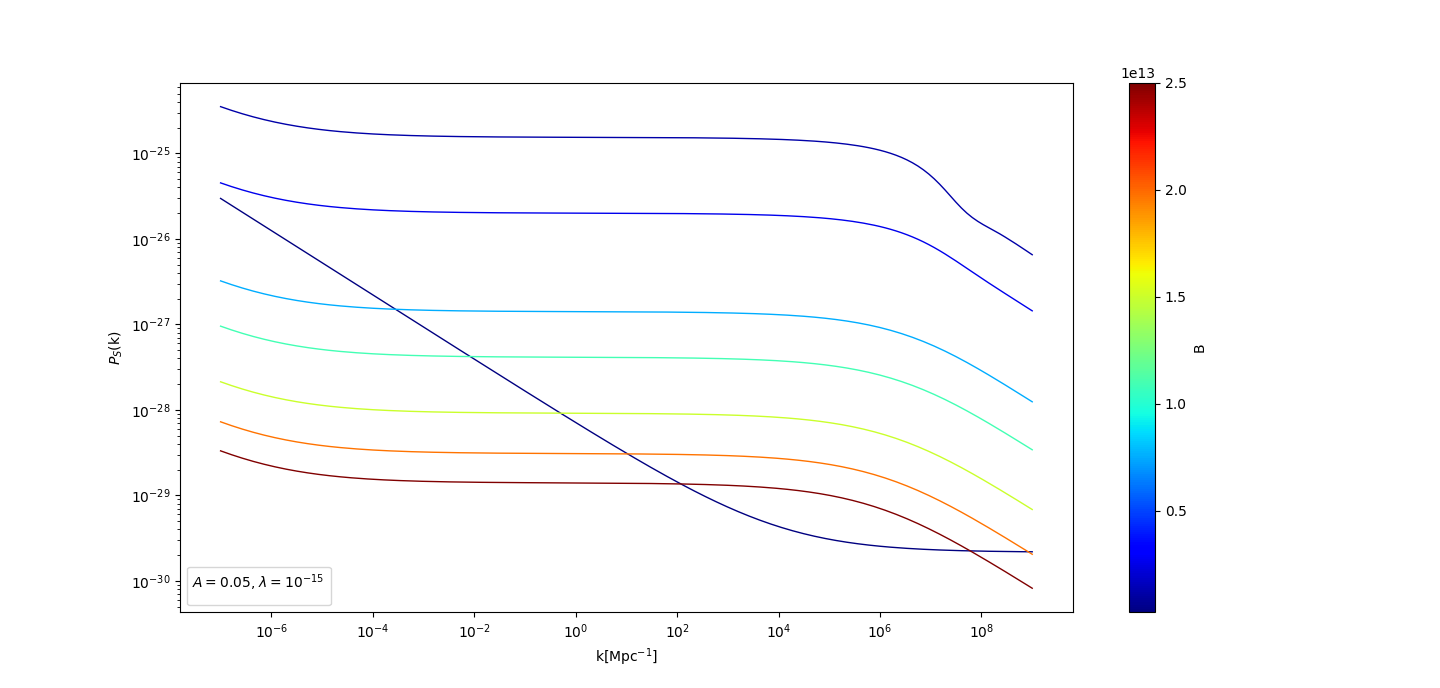}
    \includegraphics[width=0.5\textwidth]{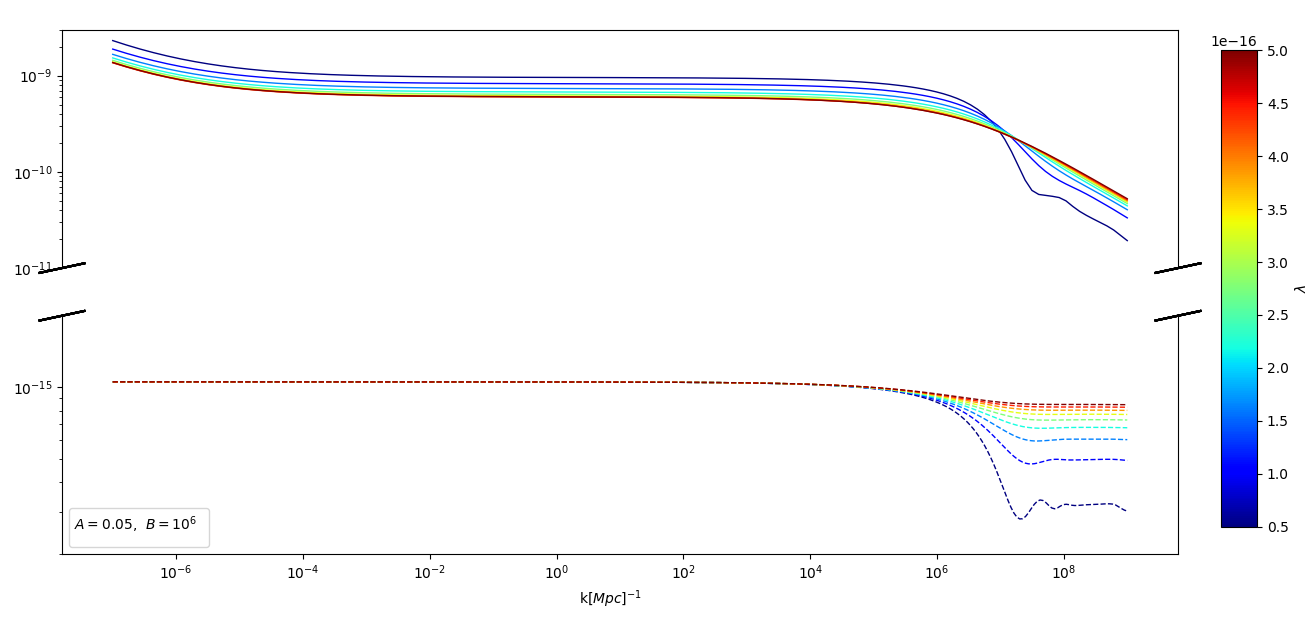}
    \includegraphics[width=0.5\textwidth]{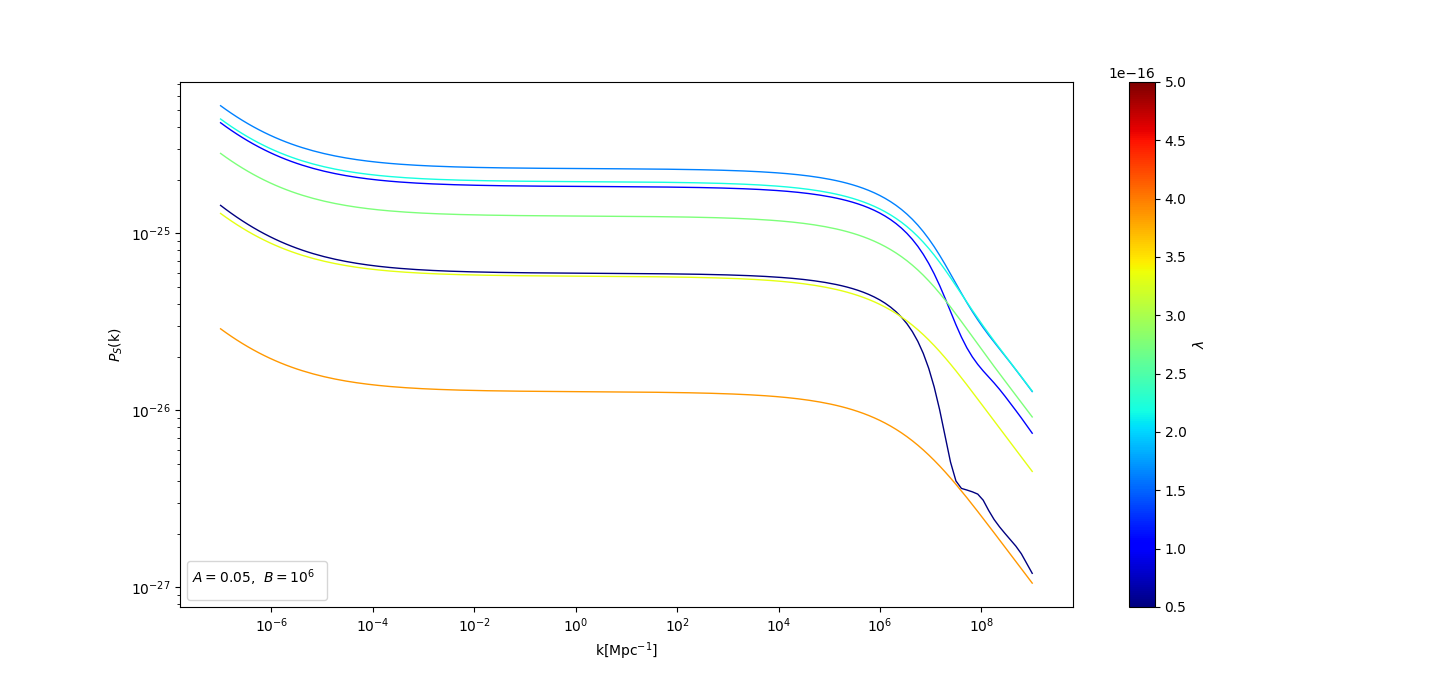}
    \caption{[Left] The curvature (solid) and tensor (dashed) power spectrum with the corresponding isocurvature power spectrum [Right], evaluated at the end of inflation. We have varied parameters, $A$ (top), $B$ (middle) and $\lambda$ (bottom) and used the same initial conditions used in \cref{fig background} }
    \label{fig power}
\end{figure}

% --------------------------------------------------------------------------------------------------------------------------

% Firstly explain what the fields are doing backed up with the potential

% Explain why there is two periods of inflation and what the importance of this is
% 

Since the three-field system is too complicated to find useful analytical solutions, we study the inflationary dynamics and the evolution of perturbations numerically. The evolution of the fields is governed by the following set of (background) equations:
\begin{eqnarray}
\ddot \psi &+& 3H\dot\psi + V_{{\rm T},\psi} = - \frac{\beta\kappa}{2}e^{-\beta\kappa\psi} \left(\dot\chi^2  + \dot\sigma^2\right),\\
\ddot \chi &+& (3H -\beta\kappa\dot\psi)\dot\chi + V_{{\rm T},\chi} = V_{{\rm T},\chi}e^{-\beta\kappa\psi},\\
\ddot \sigma &+& (3H -\beta\kappa\dot\psi)\dot\sigma + V_{{\rm T},\sigma} = V_{{\rm T},\sigma}e^{-\beta\kappa\psi}.
\end{eqnarray}

During inflation, the Noether current implies that eq. (\ref{eq:noether2}) is fulfilled after a few e-folds. Using this equation to relate $\sigma$ to $\chi$ and $\psi$ and plugging this into the potential $V_{\rm T}$, we obtain an effective two--field potential. The precise form of this effective two-field potential depends on the initial conditions for the fields, encoded in the constant ${\cal C}$ in eq. (\ref{eq:noether2}), but there are some general features, as seen in \cref{fig new potential}: in the $\psi$--direction we have plateau-like potential with a barrier near $\psi=0$, typical for $R^2$ theories. However, the addition of $\sigma$ creates a well in the potential. At the bottom of the well the potential energy vanishes. 

The dynamics of the fields depends on the initial conditions. Here, we concentrate at the case $\sigma_{\rm ini}\approx 0$ and $\chi,\psi>0$. In the first part of inflation $\sigma\approx 0$ is effectively frozen, whereas $\psi$ and $\chi$ rolling down the potential. This is a similar behaviour to that of multifield $\alpha$-attractor models \cite{Linde_attractor}. The fields settle in a local valley (shown in \cref{fig new potential}) at which point the slow--roll conditions are briefly violated, as it can be seen from the evolution of the slow-roll parameter in \cref{fig background}. In fact the condition $\eta \ll 1$ is severely violated at this point, creating additional features in the power spectrum as we will see later. From this point onward, the field $\psi$ is effectively frozen. However, $\chi$ continues to roll slowly, influenced by its coupling to $\sigma$. The field $\sigma$ evolves during this time too. Eventually $\sigma$ catches up with $\chi$ forcing the fields to fall into the global minimum, driving the potential to zero and ending inflation. The final values of $\chi$, $\sigma$ and $\psi$ are dictated by the initial conditions. 

% Briefly explain the power spectrum plots, what is seen.
% Talk about three scale
% Large scale (small k) explain the features we see
% Intermediate scales explain the features, with the flat part
% Small scale (large k) explain the feature we see and why

To analyse the resulting power spectra, we have varied each of the parameters $A$, $B$ and $\lambda$ individually in \cref{fig power}, while keeping the initial conditions for the fields fixed. We will summarize our findings in the following and refer to the Appendix \cref{3fieldnumerics} for the details on the numerical procedure. In the Appendix we also derive the perturbations of the adiabatic and entropy fields for a generic three-field system as studied in this paper. From our numerical calculations we find that $A$ determines the range of $k$--values for which the power spectrum of scalar perturbations is nearly scale--independent. This is because $A$ determines the width of the valley region in the potential. A small parameter $A$ results in an earlier start of the second period of inflation (for a given set of initial conditions). On the other hand, $B$ determines the steepness of the potential into the valley region, with smaller $B$ decreasing the gradient. Finally, $\lambda$ determines the potential energy within the valley, as it can be seen from \cref{eq full_potential}. The  parameter $B$ and $\lambda$ affect the transition to the second period of inflation and therefore affect the spike in the evolution of epsilon. As such they have a large effect on the amplitude of the power spectrum for wave numbers which leave the horizon around the transition from the first to the second period of inflation. 

As mentioned, initially $\sigma$ is effectively frozen, as illustrated in \cref{fig background}. For the initial conditions used here, the initial value of $\sigma$ is small but the initial value of $\chi$ is smaller. Because the potential gradient in the $\sigma$--direction is smaller than in the $\chi$--direction, the value of $\chi$ will eventually become larger than $\sigma$, creating a brief period of strong interactions between the two fields at that time. This causes $\chi$ to oscillate as it rolls, which can be seen from the evolution of $\dot\chi$ in \cref{fig background}. We find that while $\epsilon$ is small initially, the second slow roll parameter $\eta$ becomes large during this time, as shown in \cref{fig background}.  Eventually, after about 22 e-folds the fields settle and slow--roll inflation begins (both $\epsilon \ll 1$ and $\eta \ll 1$) and this (first) period of inflation last for over 50 e-folds. The drop in amplitude in \cref{fig power} at large scales which we observe in some runs for some choice of parameter is due to the fact that those scales cross the Hubble radius during this initial period in which $\eta$ becomes large. In the case of initial conditions where $\sigma_{\rm ini}<\chi_{\rm ini}$, the violation of the slow--roll condition $\eta\ll 1$ {\it does not occur} and there is no drop in amplitude of power at small $k$--values. 

Once the slow--roll parameter are small during the first period of inflation the produced power spectrum is approximately flat, as expected. The three fields evolve only very slowly during this period. In our numerical runs, we have chosen the parameter such that the amplitude of the scalar perturbations match data from the CMB. For large $k$--values we see a drop in the power spectrum, caused by the end of the second period of inflation as the fields fall into the global minimum of the potential. The $\sigma$--field is evolving considerably during the second period of inflation. This in turn forces the fields into a steep potential, increasing the velocity of the fields. This drop of power at small scales means our model does not predict the formation of primordial black holes. Unlike similar models to ours, our potential does not create a large entropy perturbations, we actually see the opposite effect due to the evolution of $\sigma$ from zero (this effect has been studied in similar models to ours in \cite{scalar_higgs_2018} and  \cite{Gundhi_2020_R^2_turnrate}).

% Describe the effect of varying the A what this means why we do this

% - current cosmological constraints n_s , r
% \begin{figure}
%     \centering
%     \includegraphics[scale=0.35]{n_s.png}
%     \caption{Zoomed in plot of the [Top left] tensor-to-scalar ratio, [Bottom left] scalar spectral index, [Top right] running of the spectral index, and [Bottom right] running of the running of the spectral index. Here we have used the same inital conditions and parameters as \cref{fig background}.}
%     \label{fig r ns}
% \end{figure}

The predicted power spectra are not well described by a power law. Even if we allow for a running of the spectral index $\alpha_s={\rm d}n_s/{\rm d} \ln k$ and a running of the running $\beta_s={\rm d}^2n_s/{\rm d} \ln k^2$, the usual approximation for the power spectrum $$P(k) \propto \left(\frac{k}{k_0}\right)^{n_s - 1 + \frac{1}{2}\alpha_s\ln(k/k_0) + \frac{1}{6}\beta_s (\ln(k/k_0))^2},$$ where both $\alpha_s$ and $\beta_s$ are evaluated at the pivot point $k_0$, is not a good one as both $\alpha_s$ and $\beta_s$ csn vary substantially as a function of wave-number in the model discussed. From our numerical simulations we find that the predicted tensor-to-scalar ratio $r$ is of order $10^{-6}$, substantially reduced to the two field case discussed in Section 2 (and see \cite{Ferreira_scale_independent_2019}). Other $R^2$ scale invariant models find a similar reduction in $r$, see e.g. \cite{Antoniadis_single_R2}\cite{Gialamas_single_R2}. For the parameters values $A=0.055$, $B=10^6$, and $\lambda=10^{-15}$, we find that the spectral index is consistent with current CMB observations, $n_s(k_0)\simeq 0.97$, and has a substantial running ($\alpha_s \approx 10^{-2}$) at large scales but varies widely and becomes negative at large wavenumbers. The running of the running is $\beta_s \approx -10^{-3}$ and roughly of similar order of magnitude as the running (but negative for all values of $k$). The model is an example in which isocurvature modes can cause the running of the spectral index and the running of the running to be of similar order of magnitude  \cite{vandeBruck:2016rfv}. We expect this model to be tightly constrained by current observations, but we leave this analysis for future work.

\begin{figure}
    \centering
    \includegraphics[scale=0.5]{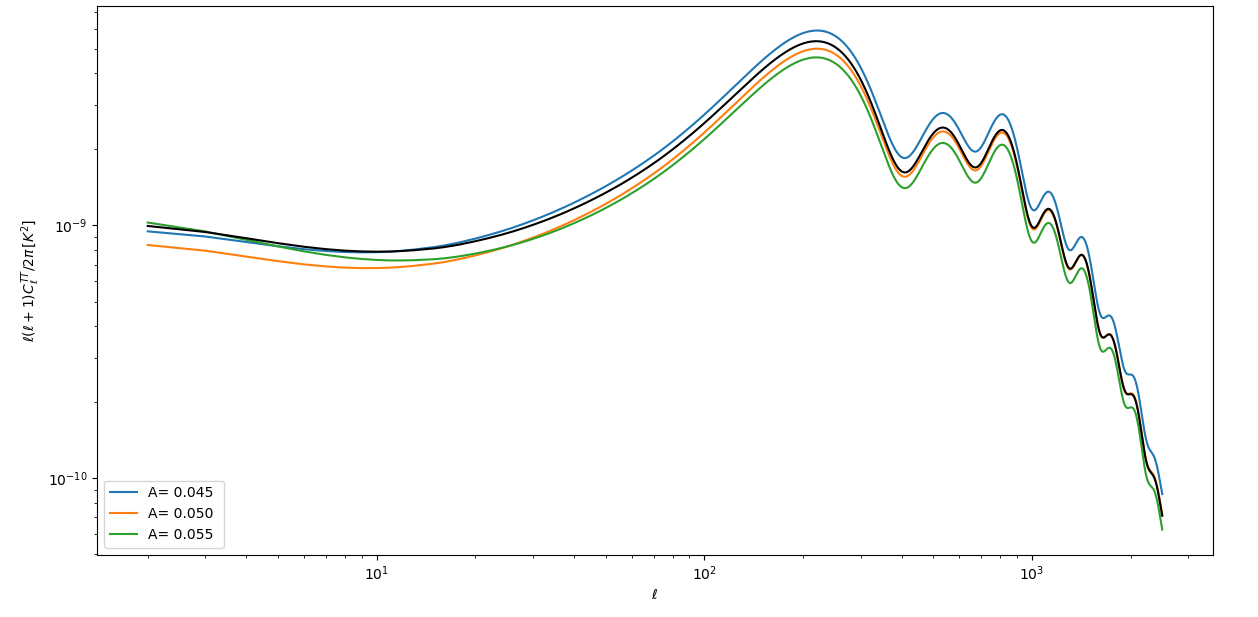}
    \caption{The CMB temperature lensed angular spectra based off our model with $B^2=5\times10^{11}$ and $\lambda=10^{-15}$, and initial conditions used in \cref{fig background}. The solid black is calculated from the standard power-law with $P_{\cal R}(k=10^{-2})= 2\times10^{-9}$ and $n_s=0.965$ for reference. }
    \label{fig c_ells}
\end{figure}

Finally we look at how the features alter the predictions of the CMB angular spectra. In \cref{fig power} is clear that a change of $B$ or $\lambda$ corresponds to a change in curvature's amplitude. As such, we have constrained the value of $B$ and $\lambda$ to fix $P_{\cal R}(k_0)\simeq 2\times10^{-9}$ with $A=0.05$. We then computed the CMB temperature angular spectra using CAMB \cite{Camb_Lewis}. We see that the features naturally manifest themselves at lower multipoles. However, even a small change in $A$ causes a shift in the amplitude of the power spectrum corresponding to a shift in amplitude in \cref{fig c_ells}.

\begin{figure}
    \centering
    \hspace{-2.6cm}\subfloat{{\includegraphics[scale=0.25]{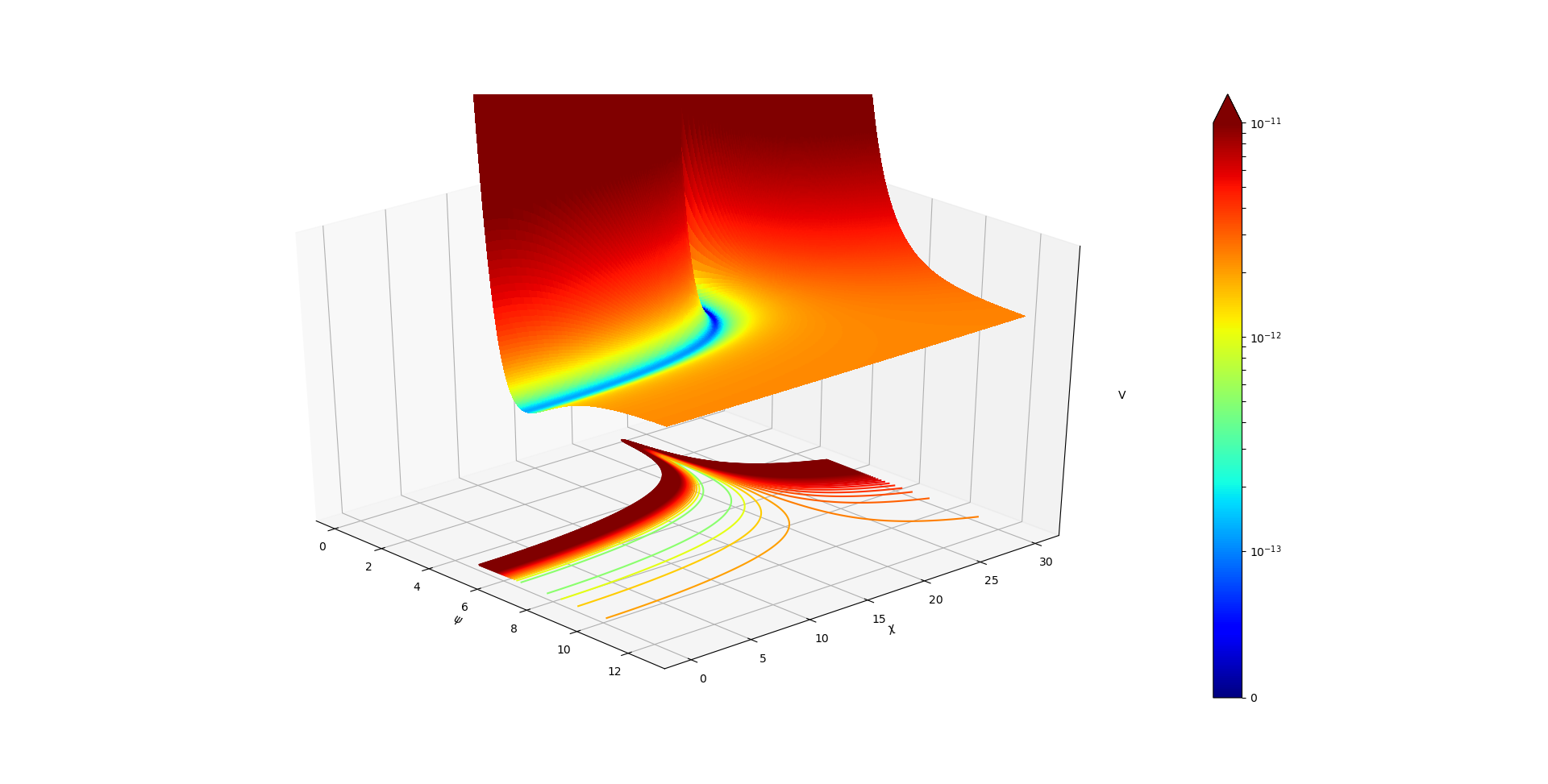} }}
    \subfloat{{\includegraphics[width=0.45\textwidth]{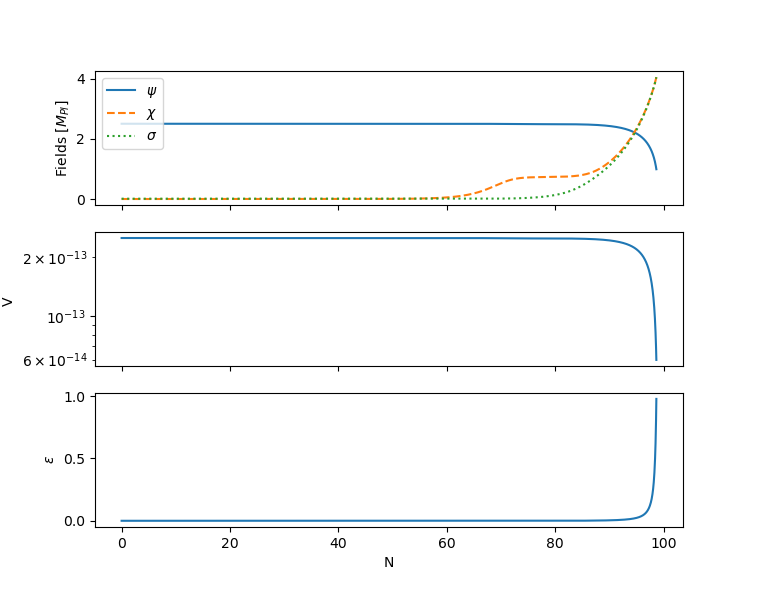} }}\qquad
    \includegraphics[width=0.6\textwidth]{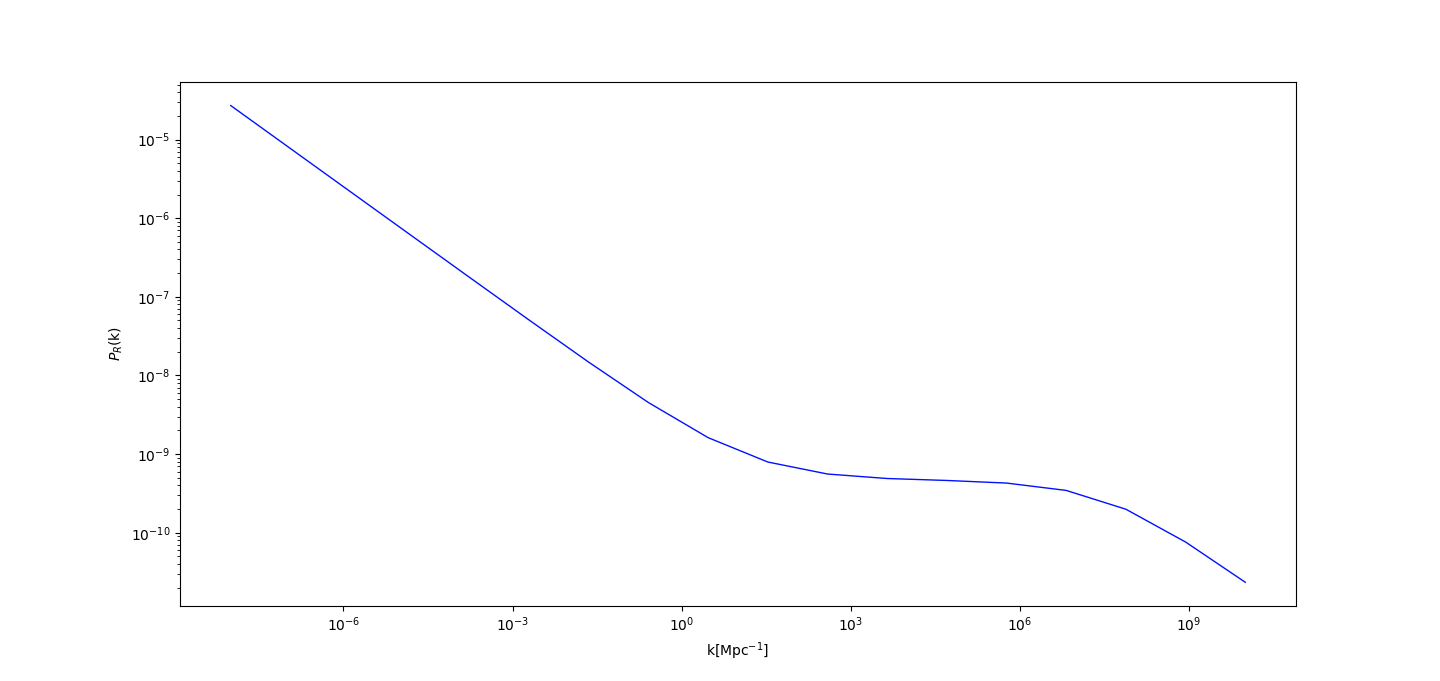}
    \caption{[Top left]Surface plot of the potential, [Top right] background fields, and [Bottom] curvature power spectrum, with $\lambda>A/B^2$. We have shown only one range of parameter, but it follows a similar trend as before with a delayed power spectrum. We used $\lambda=10^{-12}$, $A=0.05$, $B=10^{-6}$, and with the same initial conditions as \cref{fig background}}
    \label{fig lamb>AB}
\end{figure}

So far we have concentrated on the case $\lambda < A/B^2$. To understand the parameter space of our model better, we briefly discuss the case $\lambda>A/B^2$. A typical form of the potential is shown in \cref{fig lamb>AB}. We can see that this results in an effective single period of inflation with two fields, as $\chi$ and $\sigma$ become equal well before the end of inflation, tracing each others trajectory illustrated in \cref{fig lamb>AB}. There is almost a hint of another local minimum, within the potential tench of \cref{fig lamb>AB}. However, the fields will not settle here, as it can be seen from the $\chi$ trajectory. The resulting power spectrum is also shown in \cref{fig lamb>AB}. For the parameter chosen is does not fit the data at all, but we find generally that the flat part of the spectrum does not range over many orders of magnitude in $k$--values. The case $\lambda>A/B^2$ will be even more constrained than the case $\lambda<A/B^2$.   

\section{Discussion and conclusions}
We have studied the scale invariant extension of $R^2$ inflation introduced in \cite{Tambalo:2016eqr,Ferreira_scale_independent_2019}, pointing out that while it is a successful inflationary model, a non-zero cosmological constant is generated due to the self-coupling of the scalar field $\chi$, whose VEV generates the Planck mass. To remove the potentially large cosmological constant at the end of inflation, we extend the model and add another scalar field $\sigma$, with interacts with $\chi$ in a scale--invariant way, forcing the vacuum energy to vanish at the global minimum. 

We analyse the system in the Einstein frame, in which inflation is driven by three scalar fields ($\chi, \sigma$ and the scalaron $\psi$). To calculate the power spectra for scalar and tensor perturbations, we extended the two--field formalism of \cite{Wands_2000_perturbations} and \cite{Lalak_pertubations} to three fields. We used a similar methodology, creating two orthogonal components (equivalent to the entropy components), with respect to the tangential/trajectory direction in field space (corresponding to the curvature component, the details can be found in the Appendix). 

Our results show that the model can exhibit two periods of inflation. The two periods are separated by a brief (and mild) violation of the smallness of the $\epsilon$ slow roll parameter. We have numerically solved the perturbation equations and determine the power spectrum for different parameter. From the amplitude of CMB anisotropies we find that small values for the parameter $A$ are preferred ($A\approx 10^{-2}$). We identify three regimes in the predicted scalar power spectrum: at large scales, where the scales cross the horizon before all three fields begin to slow roll, the power decreases due to violation of the smallness of the $\eta$ parameter; at intermediate scales the power spectrum is flat and can match the CMB data; and at large wavenumber we find a further decrease in the amplitude for very large $k$--values (the scales leave the horizon in the second period of inflation, at the time when $\sigma$ rolls significantly and is driving the potential to zero and eventually ending inflation). 

We also discussed the predictions for the spectral index $n_s$, its running $\alpha_s$, the running of the running $\beta_s$ and the tensor--to--scalar ratio $r$. We see a substantial reduction in $r$ compared to ordinary $R^2$--inflation and its scale--invariant version of \cite{Tambalo:2016eqr,Ferreira_scale_independent_2019}. We find that the model can potentially break the usual hierarchy $n_s>|\alpha_s|>|\beta_s|$, due to entropy perturbations produced by the $\sigma$ field, making the model testable with current and future cosmological data. 

It will be interesting to study reheating and preheating in this model, given that the three fields are all coupled together. Another direction to take is to analyse this model further by studying a variety of initial conditions for the fields and constraining the model parameter with current cosmological data. 

\begin{acknowledgments}
CvdB is supported (in part) by the Lancaster--Manchester--Sheffield Consortium for Fundamental Physics under STFC grant: ST/T001038/1. RD is supported by a STFC studentship. 
\end{acknowledgments}

\appendix
\section{Three field numerical procedure}\label{3fieldnumerics}

\subsection{Linear Perturbations}
In this Appendix we derive the linear perturbation equations for the theory presented in Section III. We shall be following the work of \cite{Wands_2000_perturbations},  \cite{Finelli_2field_perturbation} and \cite{Lalak_pertubations} for the two field case and extending it to three fields. Detailed literature on the theory of cosmological perturbation can be found \cite{Brandenburger_theory_of_cosmo_pertubations}.  Working in the longitudinal gauge and with the absence of the anistropic stress, the metric is 
\begin{equation}
    ds^2= -(1+2\Phi)dt^2+a^2(1-2\Phi)d\mathbf{x}^2,
\end{equation}
where $\Phi$ is the metric perturbation in this gauge. 
The fields are broken down into background and perturbations parts,
\begin{equation}
    \psi \rightarrow \psi + \delta\psi, \hspace{0.5cm} \chi\rightarrow \chi + \delta\chi, \hspace{0.5cm} \sigma\rightarrow \sigma + \delta\sigma.
\end{equation}

This allows us to find the perturbed Klein-Gordon equations (we define $\beta\kappa\psi=-2b$ to match the notation in the literature \cite{Finelli_2field_perturbation})  
\begin{align}
    \delta\Ddot{\psi} + 3H\delta\Dot{\psi} + &\left[\frac{k^2}{a^2}  + V_{\psi\psi} -{2}b_\psi^2(\Dot{\chi}^2+ \Dot{\sigma}^2)  e^{2b}    \right]
    \delta\psi\nonumber
    \\
    & 
    + V_{\psi\chi}\delta\chi + V_{\psi\sigma}\delta\sigma -2 b_\psi e^{2b}(\Dot{\chi}\delta\Dot{\chi}+ \Dot{\sigma}\delta\Dot{\sigma}) = 4\Dot{\Phi}\Dot{\psi}- 2V_{\psi} \Phi,
    \label{eq pert KG psi}
\end{align}
\begin{align}
    \delta\Ddot{\chi}& + 3H\delta\Dot{\chi} + \left[\frac{k^2}{a^2} + V_{\chi\chi} e^{-2b}  \right]\delta\chi 
     -2b_\psi (\Dot{\chi}\delta\Dot{\psi} + \Dot{\psi}\delta\Dot{\chi}) \nonumber
     \\
     &
     +  \left(V_{\chi\psi} -2b_\psi V_\chi \right)e^{-2b}\delta\psi 
     + V_{\chi\sigma}e^{-2b}\delta\sigma
    = 4\Dot{\Phi}\Dot{\chi}- 2V_{\chi}e^{-2b}\Phi,
    \label{eq pert KG chi}
\end{align}
\begin{align}
    \delta\Ddot{\sigma}& + 3H\delta\Dot{\sigma} + \left[\frac{k^2}{a^2} + V_{\sigma\sigma}e^{-2b}   \right]\delta\sigma 
     -2b_\psi (\Dot{\sigma}\delta\Dot{\psi} + \Dot{\psi}\delta\Dot{\sigma}) \nonumber
     \\
     &
     + \left( V_{\sigma\psi}    -2b_\psi V_\sigma\right)e^{-2b}\delta\psi + V_{\sigma\chi}e^{-2b}\delta\chi 
    = 4\Dot{\Phi}\Dot{\sigma}- 2V_{\sigma}e^{-2b}\Phi.
    \label{eq pert KG sig}
\end{align}
The right hand side of the perturbed Klein-Gordon contain the metric perturbation, which is subject to the perturbed Einstein equations \cite{Brandenburger_theory_of_cosmo_pertubations} 
\begin{equation}
    3H(\Dot{\Phi} +H\Phi) + \Dot{H}\Phi + \frac{k^2}{a^2}\Phi = \frac{-\kappa^2}{2} \left[\Dot{\psi}\Dot{\delta\psi} + e^{2b}\left(\Dot{\chi}\Dot{\delta\chi} + \Dot{\sigma}\Dot{\delta\sigma} + b_\psi\delta\psi(\Dot{\chi}^2+ \Dot{\sigma}^2) \right)   + V_\psi\delta\psi + V_\chi\delta\chi + V_\sigma\delta\sigma \right],
    \label{eq energy constraint}
\end{equation}
\begin{equation}
    \Dot{\Phi} + H\Phi = \frac{\kappa^2}{2}\left[\Dot{\psi}{\delta\psi} + e^{2b}\left(\Dot{\chi}{\delta\chi} + \Dot{\sigma}{\delta\sigma}  \right) \right].
    \label{eq momentum constraint}
\end{equation}

We define the gauge invariant Mukhanov-Sasaki variables to further study the evolution of the perturbations, 
\begin{equation}
    Q_\psi=\delta\psi + \frac{\Dot{\psi}}{H}\Phi, \hspace{0.5cm} Q_\chi=\delta\chi + \frac{\Dot{\chi}}{H}\Phi, \hspace{0.5cm } Q_\sigma=\delta\sigma + \frac{\Dot{\sigma}}{H}\Phi. 
\end{equation}
Using \eqref{eq energy constraint} and \eqref{eq momentum constraint}, this allows us to rewrite our perturbed Klein-Gordon equations \eqref{eq pert KG psi} - \eqref{eq pert KG sig} as
\begin{equation}
    \ddot{Q}_\psi+ 3H(\Dot{Q}_\psi) - 2 b_\psi e^{2b\psi}\left(\Dot{\chi}\Dot{Q}_\chi+ \Dot{\sigma}\Dot{Q}_\sigma\right) + \left(\frac{k^2}{a^2} + C_{\psi\psi}\right) Q_\psi + C_{\psi\chi}Q_\chi + C_{\psi\sigma}Q_\sigma = 0,
    \label{eq pert Q psi}
\end{equation}
\begin{equation}
    \ddot{Q}_\chi + 3H\Dot{Q}_\psi 
    +2b_\psi\Dot{\chi}\Dot{Q}_\psi 
    + 2b_\psi\Dot{\psi}\Dot{Q}_\chi + \left(\frac{k^2}{a^2} + C_{\chi\chi}\right) Q_\chi + C_{\chi\psi}Q_\psi + C_{\chi\sigma}Q_\sigma = 0,
    \label{eq pert Q chi}
\end{equation}
\begin{equation}
    \ddot{Q}_\sigma + 3H\Dot{Q}_\psi 
    +2b_\psi\Dot{\sigma}\Dot{Q}_\psi 
    + 2b_\psi\Dot{\psi}\Dot{Q}_\sigma + \left(\frac{k^2}{a^2} + C_{\sigma\sigma}\right) Q_\sigma + C_{\sigma\psi}Q_\psi + C_{\sigma\chi}Q_\chi = 0,
    \label{eq pert Q sig}
\end{equation}

with the coefficients

\begin{equation*}
    C_{\psi\psi}=V_{\psi\psi}+\frac{2\kappa^2}{H}V_\psi\Dot{\psi} +  3\kappa^2\Dot{\psi}^2   -2b_\psi^2\kappa^2(\Dot{\chi}^2+ \Dot{\sigma}^2)  e^{2b}  - \frac{\kappa^4 \Dot{\psi}^2 }{2H^2}\left[\Dot{\psi}^2 + e^{2b}\left( \Dot{\chi}^2+ \Dot{\sigma}^2\right)\right]     
\end{equation*}
\begin{equation*}
    C_{\psi\rho}= V_{\psi\rho}  +\frac{\kappa^2}{H}V_\psi \Dot{\rho} e^{2b} + \frac{\kappa^2}{H}V_\rho\Dot{\psi}
    + 3\kappa^2 e^{2b}\dot{\psi}\Dot{\rho} 
    - \frac{\kappa^4 \dot{\psi}\Dot{\rho}e^{2b}}{2H^2} \left[ \Dot{\psi}^2 + e^{2b}\left( \Dot{\chi}^2+ \Dot{\sigma}^2\right)\right]
\end{equation*}
% \begin{equation*}
%     C_{\psi\sigma} = V_{\psi\sigma}  +\frac{\kappa^2}{H}V_\psi \Dot{\sigma} e^{2b} + \frac{\kappa^2}{H}V_\sigma\Dot{\psi}
%     + 3\kappa^2 e^{2b}\dot{\psi}\Dot{\sigma} 
%     - \frac{\kappa^4 \dot{\psi}\Dot{\sigma}e^{2b}}{2H^2} \left[ \Dot{\psi}^2 + e^{2b}\left( \Dot{\chi}^2+ \Dot{\sigma}^2\right)\right]
% \end{equation*}

\begin{equation*}
    C_{\rho\rho} = V_{\rho\rho}e^{-2b} 
+ \frac{2\kappa^2}{H} V_\rho\dot{\rho} +3\kappa^2\dot{\rho}^2 e^{2b}
- \frac{\kappa^4\dot{\rho}e^{2b}}{2H^2}\left[\Dot{\psi}^2 + e^{-\beta\kappa\psi}\left( \Dot{\chi}^2+ \Dot{\sigma}^2\right)\right]
\end{equation*}
\begin{equation*}
    C_{\rho\psi} = V_{\rho\psi}e^{-2b} + \frac{\kappa^2}{H}V_\psi\dot{\rho} + \frac{\kappa^2}{H}V_\rho \dot{\psi}e^{-2b} - 2b_\psi V_\rho e^{-2b} + 3\kappa^2\dot{\psi}\dot{\rho} - \frac{\kappa^4\dot{\psi}\dot{\rho}}{2H^2}\left[\Dot{\psi}^2 + e^{-\beta\kappa\psi}\left( \Dot{\chi}^2+ \Dot{\sigma}^2\right)\right]
\end{equation*}
\begin{equation*}
    C_{\rho\Tilde{\rho}} =V_{\rho\Tilde{\rho}}e^{-2b} + \frac{\kappa^2}{H} V_{\Tilde{\rho}} \dot{\rho} + \frac{\kappa^2}{H} V_{\rho}\dot{\Tilde{\rho}} + 3 {\dot{\rho}\Dot{\Tilde{\rho}}}{\kappa^2}e^{2b} - \frac{\kappa^4\dot{\rho}\dot{\Tilde{\rho}}e^{2b}}{2H^2}\left[\Dot{\psi}^2 + e^{-\beta\kappa\psi}\left( \Dot{\chi}^2+ \Dot{\sigma}^2\right)\right]
\end{equation*}

Due to the symmetry between $\chi$ and $\sigma$, as expected the perturbed Klein-Gordon equations also carry this symmetry. For brevity, we have labelled the coefficients with $\rho$, where $\rho=\chi \text{ or }\sigma$ and $\Tilde{\rho}$ is the other field.

\subsection{Tangential and normal perturbations}
For multifield inflation we break down the perturbations into tangential and normal components with respect to the trajectory of the perturbations, first introduced in \cite{Wands_2000_perturbations} and extended to nonlinear perturbations in \cite{Langlois_nonlin_pert}. The tangential perturbation equates to the curvature or adiabatic component while the normal perturbations are the entropy components. As described in \cite{Tegmark_multifield} we have one curvature component and $D-1$ entropy components, where $D$ is the number of effective fields. 

We define the curvature perturbation in the standard way \cite{Brandenburger_theory_of_cosmo_pertubations} 
\begin{equation}
    {\cal R} = \Psi - \frac{H}{\dot{H}}\left( \dot{\Psi}+ H\Phi  \right).
\end{equation}
For our $D=3$ case we find 
\begin{equation}
    {\cal R} = H\frac{\dot{\psi} Q_\psi + e^b \dot{\chi}(e^bQ_\chi) + e^b\dot{\sigma}(e^bQ_\sigma) }{\dot{\Sigma}^{ 2}} = H\frac{Q_\Sigma}{\dot{\Sigma}}. 
\end{equation}
Here, $\dot{\Sigma}^2 = \dot{\psi}^2+ e^{2b}\left(\dot{\chi}^2+\dot{\sigma}^2\right)$ is the three field adiabatic component, representing the velocity parallel to the trajectory. This allows us to define $Q_\Sigma$ as the instantaneous curvature Mukhanov-Sasaki variable corresponding to the perturbations parallel to the trajectory $(Q_\psi,e^bQ_\chi ,e^b Q_\sigma)$, defined in the same way as \cite{Wands_2000_perturbations}.
Following the methodology laid out in \cite{Lalak_pertubations}, we set up our three field model by extending the 2D circular coordinate system to a 3D spherical system, using $Q_\Sigma$ as an effective radial component,
\begin{equation}
    Q_\Sigma = \sin\varphi\left(\cos\theta Q_\psi + \sin\theta e^bQ_\chi \right) + \cos\varphi e^b Q_\sigma.
\end{equation}
After equating coefficients and some basic algebra we determine, 
\begin{align}
    \cos(\varphi) = \frac{e^b \dot{\sigma}}{\dot{\Sigma}}; 
    \hspace{1cm} 
    \sin(\varphi) = \frac{\sqrt{\dot{\psi}^2 + e^{2b}\dot{\chi}^2}}{\dot{\Sigma}}; 
    \hspace{0.5cm}
    \cos(\theta) = \frac{\dot{\psi}}{\sqrt{\dot{\psi}^2 + e^{2b}\dot{\chi}^2}};
    \hspace{0.5cm}
    \sin(\theta) = \frac{e^b\dot{\chi}}{\sqrt{\dot{\psi}^2 + e^{2b}\dot{\chi}^2}}
    \label{eq angle}
\end{align}

Now with our angles established we can calculate the entropy perturbations, $\delta s_1$ and $\delta s_2$, which are defined to be orthogonal to the trajectory, $Q_\Sigma$: 

\begin{equation}
    \delta s_1 = \cos(\varphi)\left[\cos(\theta)Q_\psi + \sin(\theta)(e^b Q_\chi)  \right] - \sin(\varphi) (e^b Q_\sigma), \hspace{0.75cm} \delta s_2 = - \sin(\theta)Q_\psi + \cos(\theta)(e^b Q_\chi).
\end{equation}
Using \eqref{eq angle} we can rewrite this in terms of the fields, 
\begin{equation}
    \delta s_1 = \frac{e^b\dot{\sigma}}{\dot{\Sigma}\sqrt{\dot{\psi}^2 + e^{2b}\dot{\chi}^2}}\left[\dot{\psi}Q_\psi + e^b\dot{\chi}(e^b Q_\chi) \right] - \frac{\sqrt{\dot{\psi}^2 + e^{2b}\dot{\chi}^2}}{\dot{\Sigma}}(e^b Q_\sigma), \hspace{0.75cm} \delta s_2 = \frac{-e^b\dot{\chi} Q_\psi + \dot{\psi}(e^b Q_\chi)}{\sqrt{\dot{\psi}^2 + e^{2b}\dot{\chi}^2}}.
\end{equation}
We can then calculate the entropy component as ${\cal S}_i =\delta s_i H/\dot{\Sigma}$. More importantly, we now have a complete system for our three field system that allows us to relate our field perturbations into tangential and normal components via the rotational matrix, 
\begin{equation}
    \begin{pmatrix}
    Q_\Sigma\\\delta s_1\\ \delta s_2
    \end{pmatrix}
    =
    \begin{pmatrix}
    \sin(\varphi)\cos(\theta) & \sin(\varphi)\sin(\theta)& 
    \cos(\varphi)\\
    \cos(\varphi)\cos(\theta) & \cos(\varphi)\sin(\theta)& 
    -\sin(\varphi)\\
    -\sin(\theta)& \cos(\theta)&0
    \end{pmatrix}
    \begin{pmatrix}
    Q_\psi\\e^b Q_\chi\\e^b Q_\sigma
    \end{pmatrix}.
    \label{eq matrix}
\end{equation}

Finally, we can determine how the curvature component is sourced. To do this, and for completion, we find the Klein-Gordon equation for our adiabatic field, 
\begin{equation}
    \ddot{\Sigma} + 3H\dot{\Sigma} + V_\Sigma = 0,
\end{equation}
and the turning rate, 
\begin{equation}
    \dot{\varphi} = -\frac{V_{s_1}}{\dot{\Sigma}} + b_\psi \dot{\Sigma} \cos\theta \cos\varphi, \hspace{0.75cm} \dot{\theta} = - \frac{V_{s_2}}{\sqrt{\dot{\psi}^2 + e^{2b}\dot{\chi}^2}} - \frac{\dot{\Sigma}}{\sqrt{\dot{\psi}^2 + e^{2b}\dot{\chi}^2}}b_\psi\dot{\Sigma}\sin\theta.
\end{equation}
The adiabatic and entropy potentials follow the same rotation as \eqref{eq matrix},
\begin{equation}
    \begin{pmatrix}
    V_\Sigma\\V_{s_1}\\V_{s_2}
    \end{pmatrix}
    =
    \begin{pmatrix}
    \sin(\varphi)\cos(\theta) & \sin(\varphi)\sin(\theta)& 
    \cos(\varphi)\\
    \cos(\varphi)\cos(\theta) & \cos(\varphi)\sin(\theta)& 
    -\sin(\varphi)\\
    -\sin(\theta)& \cos(\theta)&0
    \end{pmatrix}
    \begin{pmatrix}
    V_\psi\\e^{-b} V_\chi\\e^{-b} V_\sigma
    \end{pmatrix}.
\end{equation}
Taking the derivative of $\cal R$ with respect to time we find a nice compact version, similar to that of \cite{Wands_2000_perturbations},  \cite{Finelli_2field_perturbation}, and \cite{Finelli_2020_second_pap}
\begin{equation}
    \dot{{\cal R}} = \frac{H}{\dot{H}}\frac{k^2}{a^2}\Phi - 2\frac{H}{\dot{\Sigma}^2}(V_{s_1}\delta s_1 + V_{s_2}\delta s_2).
    \label{eq source}
\end{equation}
In the derivation, we have used the comoving energy density found from the energy and momentum constraints (\eqref{eq energy constraint} and \eqref{eq momentum constraint}), 
\begin{equation}
    \epsilon_m= -\frac{2}{\kappa^2}\frac{k^2}{a^2}\Phi =
    \dot{\Sigma} \dot{Q}_\Sigma + V_\Sigma Q_\Sigma + 2V_{s_1}\delta s_1 + 2V_{s_2}\delta s_2 + \left( 3H + \frac{\dot{H}}{H}\right)\dot{\Sigma}Q_\Sigma.
\end{equation}

\subsection{Numerical setup}\label{numerical_setup}
For the numerical calculations, we work with the more convenient time variable $N$ (the number of e-folds, defined by $dN = H dt$) rather than cosmic time. We also set the pivot scale $k_*=0.05\text{ Mpc}^{-1}$ to leave the Hubble radius 50 e-folds before the end of inflation. We impose Bunch-Davis initial conditions (BD) on our tangential and normal perturbations ($Q_\Sigma, \delta s_1, \delta s_2$), separately, and deep inside the Hubble radius. To do this, we integrate \eqref{eq pert Q psi}-\eqref{eq pert Q sig}, with the initial conditions, 
\begin{equation}
    \begin{pmatrix}
    Q_\psi\\e^{b} Q_\chi\\e^{b} Q_\sigma
    \end{pmatrix}
    =
    \begin{pmatrix}
    \sin(\varphi)\cos(\theta) & \cos(\varphi)\cos(\theta) & 
    -\sin(\varphi)\\
     \sin(\varphi)\sin(\theta)& \cos(\varphi)\sin(\theta)& 
    \cos(\theta)\\
    -\cos(\varphi)& -\sin(\varphi)&0
    \end{pmatrix}
    \begin{pmatrix}
    Q_\Sigma\\\delta s_1\\ \delta s_2
    \end{pmatrix}.
\end{equation}
To confirm that integration starts deep inside the Hubble radius, we impose these initial conditions when $k = 50 aH$.

To ensure that there is no correlation between the curvature and entropy modes deep inside the Hubble radius, we integrate the perturbation equations three times, with different initial conditions. In the first run, we choose $Q_\Sigma$ to have BD initial conditions and $\delta s_1 =\delta s_2 =0$; then we permute these initial conditions on each of the components. We indicate the run by roman numerals, so that the results of each run are ${\cal R}_{\text{I}}$, ${\cal R}_{\text{II}}$ and ${\cal R}_{\rm{III}}$ and correspondingly for ${\cal S}_{1,(I,II,III)}$ and ${\cal S}_{2,(I,II,III)}$. 
This will then allow us to compute the curvature and isocurvature power spectrum from our three runs and obtain the prediction for the power spectra as follows:  

\begin{equation}
    {\cal P_R}(k) = \frac{k^3}{2\pi^2}\left(|{\cal R}_\text{I}|^2 + |{\cal R}_\text{II}|^2 + |{\cal R}_\text{III}|^2   \right)
\end{equation}
\begin{equation}
    {\cal P}_{{\cal S}_1}(k) = \frac{k^3}{2\pi^2}\left(|{\cal S}_{1,\text{I}}|^2 + |{\cal S}_{1,\text{II}}|^2 + |{\cal S}_{1,\text{III}}|^2   \right)
\end{equation}
\begin{equation}
    {\cal P}_{{\cal S}_2}(k) = \frac{k^3}{2\pi^2}\left(|{\cal S}_{2,\text{I}}|^2 + |{\cal S}_{2,\text{II}}|^2 + |{\cal S}_{2,\text{III}}|^2   \right)
\end{equation}
We compute the tensor power spectrum as follows (see e.g. \cite{Riotto_Inflation}): 
The mode equation for gravitational waves is given by 
\begin{equation}
    v_k''+\left(k^2 -\frac{a''}{a}\right)v_k = 0,
\end{equation}
where $'$ denotes the derivative with respect to conformal time.
The power spectrum is then given by 
\begin{equation}
    P_T(k)= \frac{k^3}{2\pi^2} \left|\frac{v_k}{a}\right|^2.
    \label{eq P_T}
\end{equation}

\bibliography{references.bib}

\end{document}